\documentclass[a4paper]{article}
\usepackage[ansinew]{inputenc}
\usepackage{times}
\usepackage[T1]{fontenc}
\usepackage{graphicx}
\usepackage{geometry}
\usepackage{amssymb}
\usepackage{amsmath}

\newcommand{\Robject}[1]{\texttt{#1}}
\usepackage{rotating}

\usepackage{xcolor}
\colorlet{shadecolor}{gray!25}
\usepackage{subfig}
\usepackage{float}
\usepackage{lscape}

\title{Evaluation of histological findings with severity grade, to analyze toxicology in-vivo studies}
\author{Ludwig A. Hothorn\\ 
\footnotesize (retired from) Leibniz University Hannover, Germany, ($ludwig@hothorn.de$),
\normalsize\\
Klaus Weber\\ 
\footnotesize AnaPath Services GmbH, 4625 Oberbuchsiten, Switzerland ($kweber@anapth.ch$)}
\normalsize
				
\date{\today}

\begin{document}
\maketitle
\begin{abstract}
In-vivo toxicological studies are characterized by multiple primary endpoints with quite different scales. Whereas guidelines and publications provide various statistical tests for normally distributed endpoints (such as organ weights) and proportions (such as tumor rates), few approaches are available for graded histopathological findings, such as 0, +, ++, +++. This represents a basic contradiction of the statistical analysis because these graded findings sometimes show a high predictive value for potential toxic effects. Here we discuss different methods comparatively, especially from the viewpoints of i) designs for very small sample sizes and ii) interpretability by toxicologists.
A new approach is recommended where a simultaneous test is performed over all class combinations of score levels, such as (0, +) vs (++, +++). Corresponding R code is provided by way of a data example.
\end{abstract}

\normalsize

\section{The problem}
The basic dilemma in analyzing in-vivo toxicology studies is that both methods and software are available to analyze dose-response data of in-vivo assays, including the important zero-dose control, when data (to be precise: residuals) are normal-distributed, such as body or organ weights, but their toxicological relevance is often rather limited. More relevant toxicologically are proportions, such as mortality and tumor rates, or even more so, histological findings with severity grades. For the latter, rather few methods/software are available. As a rule of thumb, one can assume a sequence of statistical precision (e.g., in terms of power):  normal $\rightarrow$ score $\rightarrow$.   But frequently the reverse order is true for the toxicological relevance. This approach may be considered old-fashioned, however it is still practiced. Regardless, the three-cross system easily translates into numerical, ordered categorical values. In toxicology, the system is used to describe, for example, the severity of clinical or ophthalmological symptoms, where the crosses or numbers represent the adverbs 'minimal', 'moderate', and 'severe'. In such cases, no definition is attributed to these values yet it is important that these values be defined. In other cases, e.g., for hematology values, attempts have been made to define the grading by a range of deviations from a mean value established by control data \cite{deKort2020}.
Otherwise, there are guidances/guidelines describing the use of three-grade (plus 0) systems, e.g. to establish thyroid gland alterations in the amphibian metamorphosis assay \cite{Grim2009}, \cite{OECD2014}, \cite{OECD2019}. In this specific case, each grade is defined by an alteration range, e.g., for thyroid atrophy: Grade 0 -  less than a $20\%$ reduction in size in comparison to controls; Grade 1 - gland size is $30\%-50\%$ reduced from the size of control glands; Grade 2 - gland size is $60\%-80\%$ reduced from the size of control glands; Grade 3 - gland size is over $80\%$ reduced from the size of control glands.  In addition, there are proposed systems based on zero and four severity degrees, where the values are quasi-defined, e.g., for the evaluation of local reactions due to the implantation of medical devices or their materials. In this case, examples are presented for the cell type and tissue reaction that are described, e.g., lymphocytes per high power field (hpf=magnification x400): grade 0: 0; grade 1: rare, 1-5 per hpf; grade 2: 5-10 per hpf; grade 3: marked infiltration; grade 4: packed.\\
Currently, a five-grade system (plus 0) is applied by almost all toxicological pathologists based on recommendations of Thoolen et al., 2010 \cite{Thoolen2010}. The gradings should however be described, hence an STP best practice paper \cite{Morton2006} on pathology report writing recommends: \textit{When severity grading is important to the understanding of major study findings, it may be useful to provide a description of the distinguishing features of each severity grade}.  In fact, an STP working group supports greater transparency and consistency in the reporting of grading scales and provided several recommendations on detailed severity criteria \cite{Schafer2018}. A published example by Weber et al. supports a graded scale, in this instance for pathological changes in the lungs \cite{Weber2018}.

Here we consider scores with only a few categories, e.g., $0, +, ++, +++$. Again, the number of informative categories matters, i.e., a three-category score is more precise than a proportion (two categories) whereas $k > 5$ category scores behave nearly like a normally distributed endpoint (whereby commonly not every individual category is informative in the respective assay). 
Several statistical approaches are available for scores or count data. The focus here is interpretability of both the effect size and its confidence interval \cite{Hothorn2016}.

\subsection{A motivating example}
As a simple motivating example, the liver basophilia severity scores data \cite{Green2014}, Table 4 was selected.
\begin{table}[ht]
\centering
\caption{c-by-k Table Data of Liver Basophilia Severity Scores} 
\label{tab:green}
\begin{tabular}{c|rrrr}
         & Severity & 1 & 2 & 3 \\ 
  Dose &            &     &     &     \\ 
   \hline
1    &            &   9 &   4 &   1 \\ 
  2    &            &   4 &   8 &   0 \\ 
  3    &            &   7 &   6 &   0 \\ 
  4    &            &   3 &  10 &   1 \\ 
  5    &            &   6 &   4 &   4 \\ 
  \end{tabular}
\end{table}

Dose=1 is the zero-dose control. Notice that no zero-severity score (0) is defined in this assay (somewhat curious in a toxicological bioassay or else the zero data have not been presented). Two questions should be answered in this context: i) is there a score change for any dose compared to the control and could this be in terms of a dose trend, and ii) is this score change with dose more likely from category 1 to 2, 2 to 3 or  1 to 3?

\section{Discussion of several approaches to analyzing scores or count data}
\subsection{Classifications and assumptions}
First, it may not seem too important but should we really use two-sided tests as proposed in some guidances and used commonly in publications? The answer is clearly no, both from a power argument (in designs with small $n_i$) and from the argument of problem-adapted interpretation (there is no interest in a possible dose-related shift from +++ to +). Appropriate and a-priori chosen one-sided tests should clearly be preferred in most scenarios in toxicological risk assessment.\\
A second question when evaluating the common design $[C; D_1,...,D_k]$ is whether to use i) (one-sided) many-to-one comparisons of the Dunnett-type \cite{Dunnett1955}, ii) trend tests assuming the doses qualitatively in comparison to C-, i.e., Williams-type \cite{Williams1971} or iii) trend tests assuming doses quantitatively, i.e., Tukey-type \cite{Schaarschmidt2021}. It may be an unconventional idea, but we recommend all three options simultaneously, i.e., answering an imprecisely formulated question for a possible dose-related toxicological effect by several models. One must pay a price in power loss for the multiplicity of such an approach. The power loss, however, is not too dramatic considering highly correlated tests and is outweighed by the advantage, broader interpretation possibility.\\
A third issue is how to transform the graded findings, $[0,+,++,+++]$, into ordered categorical data, $[0,1,2,3]$. It remains an ordinal response: 0 is less toxic than 1, etc. Several methods will be discussed. One selection aspect would by the min(p) idea, i.e., take the best. Toxicologists' reasoning is more \textit{ 'Is the trend caused by +++ shift alone, or jointly (++, +++), or...?'}. Two serious constraints are common in such toxicological data: i) very small $n_i$, so small that asymptotic methods have problems (e.g., with FWER control), and ii) rather equal low score values in the control, i.e., $p_0 = 0$ is possible (and toxicologically not surprising), and thus data without variance, which causes massive convergence problems in the general linear model (GLM) logistic.
A further limitation of the sliced Cochran-Armitage test is that it presents a test for an exactly linear trend, but in toxicology such a perfect linear dose-response relationship is quite rare in practice. Therefore, we propose several trend tests, both Tukey-type for dose as a quantitative covariate and Williams/Dunnett-type for dose as a qualitative factor for pooling patterns of the categories \cite{LH2020}.

\subsection{Using Dunnett, Williams or Tukey-type tests?}
Confusingly, for the evaluation of the widely used design $[C-, D_1, ...,D_k]$ (where $D_i$ represents the administered doses and C- is the zero-dose control), some guidelines recommend the Dunnett's test \cite{Dunnett1955}, others the Williams test \cite{Williams1971}, and occasionally, the Armitage test \cite{Armitage1955}.
In the most commonly used Dunnett's test, each dose is compared to C- (one-sided, e.g., for increase or in many cases, two-sided for any difference) with control of family-wise error rate for all these comparisons. In the Williams test, a monotone increase (or decrease) effect with increasing dose is tested in comparison to C- (usually one-sided), whereby an individual comparison to C- can only   be interpreted for the maximum dose. The Armitage test is a trend test for a linear increase (or decrease) in proportions, such as tumor rates. As a special regression model, it takes the doses into account quantitatively, whereas Dunnett's and Williams' tests consider the doses as a qualitative factor. The disadvantage   of the Armitage test, that it is only defined for linear alternatives and proportions, can be overcome  with the Tukey test \cite{Tukey1985}. This simultaneously models linear, ordinal and logarithmic doses in the generalized linear model and is thus sensitive to a wide class of possible dose-response dependencies \cite{Schaarschmidt2021}. The choice of one of these tests is nontrivial and depends on the design, the question, and the data condition \cite{LH2020}.

\subsection{A nonparametric approach based on relative effect size}
Nonparametric tests (e.g., Kruskal-Wallis or Mann-Whitney tests \cite{Yamaguchi2019}) are commonly used for ordinal response data but tied data are considered only an approximation and confidence intervals are hard to obtain or to interpret. New proposals for related trend tests, e.g., Altunkaynak and Gamgam \cite{Altunkaynak2020} suffer from the same issues. The relative effect size approach for multiple contrasts allows i) discrete data, ii) variance heterogeneity and iii) provides simultaneous confidence intervals \cite{Konietschke2012}. A related CRAN library \Robject{nparcomp} is available \cite{Konietschke2015}. Simple code for the example data, in this case for the Williams-type test \cite{Konietschke2011}, is:

\footnotesize
\begin{verbatim}
library(nparcomp)
ko1<-nparcomp(Severity ~dose, data=green, asy.method = "mult.t",
            type = "Williams",alternative = "greater", info = FALSE)
summary(ko1)  # Reveal the adjusted p-values
plot(ko1)  # Plot simultaneous confidence limits
\end{verbatim}
\normalsize

Adjusted p-values or simultaneous confidence intervals are available whereas the interpretation of relative effect size as success probability is challenging. The win odds ratio \cite{Brunner2021} may be an alternative for better understanding by toxicologists.

\footnotesize
\begin{verbatim}
  Comparison Estimator Lower Upper Statistic    p.Value
1 p( 1 , 2 )     0.631 0.398     1 1.2770958 0.26414207
2 p( 1 , 3 )     0.536 0.309     1 0.3577455 0.65530831
3 p( 1 , 4 )     0.699 0.483     1 2.0936297 0.07009149
4 p( 1 , 5 )     0.638 0.416     1 1.4123519 0.21881077
\end{verbatim}
\normalsize

\subsection{Simple transformation model}
Simple transformation of the endpoint allows the use of standard tests, such as Dunnett's or Williams'.
Here the Freeman-Tukey transformation \cite{FREEMAN1950} for count data is proposed for use in small sample designs in
toxicology \cite{Hothorn2013}. Such a simple approach is robust, particularly when using designs with small $n_i$, which
represents a serious argument. Adjusted p-values or simultaneous confidence intervals are available but
intervals are hard to interpret because back-transformation is not always obvious.

\footnotesize
\begin{verbatim}
library("multcomp")
green$FT <-sqrt(green$Severity)+ sqrt(green$Severity+1)
modFT <-lm(FT~dose, data=green)
pvalFT <-summary(glht(modFT, linfct=mcp(dose="Dunnett"),alternative="greater"))
\end{verbatim}
\normalsize
Adjusted p-values are available:
\footnotesize
\begin{verbatim}
Linear Hypotheses:
           Estimate Std. Error t value Pr(>t)  
2 - 1 <= 0  0.18475    0.17167   1.076 0.3468  
3 - 1 <= 0  0.03458    0.16808   0.206 0.7312  
4 - 1 <= 0  0.31374    0.16494   1.902 0.0943 .
5 - 1 <= 0  0.28239    0.16494   1.712 0.1341  
\end{verbatim}
\normalsize
\subsection{Generalized linear model}
The generalized linear model (GLM) is the first line natural approach for such count data, particularly using the flexible quasi-Poisson link function. However, this model is only defined asymptotically. With the object-oriented properties within R, the estimation of a GLM object can be easily imported into the function \Robject{glht()} in the \Robject{multcomp} package, based on the theory of simultaneous inference in general parametric models \cite{Hothorn2008}.

\footnotesize
\begin{verbatim}
library("multcomp")
exP <-glm(Severity~dose, data=green, family=quasipoisson(link = "log"))
pvalGLM <-summary(glht(exP, linfct=mcp(dose="Dunnett"), alternative="greater"))
\end{verbatim}
\normalsize
Adjusted p-values or simultaneous confidence intervals are available, where the effect size is the odds ratio.

\footnotesize
\begin{verbatim}
Linear Hypotheses:
           Estimate Std. Error z value Pr(>z)
2 - 1 <= 0  0.15415    0.15472   0.996  0.367
3 - 1 <= 0  0.02281    0.15674   0.146  0.739
4 - 1 <= 0  0.26236    0.14552   1.803  0.105
5 - 1 <= 0  0.26236    0.14552   1.803  0.105
\end{verbatim}
\normalsize

\subsection{Cumulative link model}
The cumulative link model is based on the assumption that an ordinal endpoint $Z_{ij}$ falls into $j = 1, \ldots, J$ categories in each of $i$ groups and follows a multinomial distribution with the parameters $\pi_{ij}$. Then the cumulative probabilities
$ Pr(Z_i \leq j) = \pi_{i1} + \ldots + \pi_{ij}$ can be defined  accordingly with
cumulative logits: $ \textup{logit}(Pr(Z_i \leq j)) =
  \log \frac{Pr(Z_i \leq j)}{1 - Pr(Z_i \leq j)}$. This framework can be used within  regression models independently for each $j$ with the package \Robject|ordinal| \cite{Christensen2020}. Particularly, using the \Robject|clm| function the parameter of a cumulative link model can be fitted whereas the endpoint \verb|severe| is defined as an ordered categorical.
The effect sizes are cumulative log-odds ratios and the parameter estimates are differences from control (abbreviated with $1$)). Therefore, Wald-type confidence intervals can be estimated in case Bonferroni adjustments for a Dunnett-type approach were used level $(1-0.05/2)$ because of four one-sided rather than two-sided comparisons.

\footnotesize
\begin{verbatim}
GreenCLM <- clm(sever ~ dose, data=green)
propCI <-exp(confint(GreenCLM, level=1-0.05/2)) # Bonferroni odds ratio CI
\end{verbatim}
\normalsize

All simultaneous lower limits are less than one, i.e., no increasing severity from category $1 \Rightarrow 2$ and $2 \Rightarrow 3$ occur in the comparison of any dose with respect to control.

\subsection{Ordered categorical regression}
The ordered categorical regression model estimates the proportional log-odds ratio for both the treatment contrast estimates and the intercepts \cite{Hothorn2020a}. Using the  \Robject{Polr} function that comes with the CRAN package \Robject{tram}, these estimates are available and can be used easily by the object properties for simultaneous inference, doses vs. control, in the function \Robject{glht}:

\footnotesize
\begin{verbatim}
library(multcomp)
library(tram)
green$score<-as.ordered(green$Severity)
PL2<-polr(score ~ dose, data =green)
summary(PL2)
summary(glht(PL2, linfct=mcp(dose="Dunnett"),
                     alternative="greater"))
\end{verbatim}
\normalsize

Adjusted p-values or simultaneous confidence intervals are available, where the effect size is the odds ratio.

\footnotesize
\begin{verbatim}
Linear Hypotheses:
           Estimate Std. Error z value Pr(>z)  
2 - 1 <= 0   0.9630     0.7743   1.244 0.2686  
3 - 1 <= 0   0.2863     0.7743   0.370 0.6510  
4 - 1 <= 0   1.4950     0.7628   1.960 0.0767 .
5 - 1 <= 0   1.3631     0.8102   1.682 0.1324  
\end{verbatim}


\subsection{Multinomial model}
A further modeling option is to define the scores as a multinomial vector \cite{Schaarschmidt2017}.  The  \Robject{VGAM} package allows fitting multinomial models by the \Robject{vglm()} function, whereas several R-functions provide simultaneous inference between both treatment contrasts and categories \cite{Vogel2018}:

\footnotesize
\begin{verbatim}
library(VGAM)
green$C1 <- ifelse(green$Severity == 1, 1, 0)
green$C2 <- ifelse(green$Severity == 2, 1, 0)
green$C3 <- ifelse(green$Severity == 3, 1, 0)
multivgam <- vglm(cbind(C1,C2,C3) ~ dose,
family=multinomial(refLevel=1), data=green)
summary(glht(model = multin2mcp(multivgam, dispersion="overall"),
linfct = mcp2matrix(multivgam, linfct = mcp(dose = "Dunnett"))$K, alternative="greater"))
@
\end{verbatim}
\normalsize
The estimated adjusted p-values reveal a non-significant effect for comparing dose 4 vs. control for the comparison of categories $C2/C1$. Notice the extreme variance estimators due to too low variance in doses 1 and 2 for $C3/C1$.

\footnotesize
\begin{verbatim}
Linear Hypotheses:
                   Estimate Std. Error t value Pr(>t)  
C2/C1: 2 - 1 <= 0    1.5041     0.8044   1.870 0.1961  
C2/C1: 3 - 1 <= 0    0.6568     0.7678   0.855 0.7302  
C2/C1: 4 - 1 <= 0    2.0149     0.8357   2.411 0.0604 .
C2/C1: 5 - 1 <= 0    0.4055     0.8269   0.490 0.8840  
C3/C1: 2 - 1 <= 0  -15.1809  1688.2201  -0.009 0.9788  
C3/C1: 3 - 1 <= 0  -15.5521  1536.4467  -0.010 0.9788  
C3/C1: 4 - 1 <= 0    1.0986     1.4659   0.749 0.7822  
C3/C1: 5 - 1 <= 0    1.7918     1.1589   1.546 0.3405  
\end{verbatim}
\normalsize

\subsection{Most likely transformation model, special for count data}
To overcome 1st moment modeling, adjusting or ignoring the other moments of the distribution allows the most likely transformation (MLT) model  \cite{Hothorn2020a}. Count data are particularly sensitive to the underlying assumption, e.g.,  for ties and
variance heterogeneity structure, and therefore a special MLT approach for counts was recently proposed  \cite{Siegfried2020}. The CRAN package \Robject{cotram} offers flexible count transformation models where count responses may arise from various and complex data-generating processes.  Again, using R's object-oriented properties, the estimate of a cotram object can be easily imported into the function \Robject{glht()} in the \Robject{multcomp} package. 

\footnotesize
\begin{verbatim}
library(cotram)
NSH<-cotram(Severity~dose, data=green,order=2)
WISH<-glht(NSH, linfct = mcp(dose = "Dunnett"), alternative="greater")
\end{verbatim}
\normalsize
Adjusted p-values or simultaneous confidence intervals are available, where the effect size is the odds ratio.
\footnotesize
\begin{verbatim}
Linear Hypotheses:
           Estimate Std. Error z value Pr(>z)  
2 - 1 <= 0   0.9815     0.7790   1.260 0.2635  
3 - 1 <= 0   0.2901     0.7765   0.374 0.6508  
4 - 1 <= 0   1.5316     0.7700   1.989 0.0723 .
5 - 1 <= 0   1.3796     0.8150   1.693 0.1304
\end{verbatim}
\normalsize  

\section{A new proposal: multiple contrast test based on multiple correlated binary data}
This new proposal based on five items. First, collapsing of certain categories to obtain a toxicologists-like interpretability \cite{Green2014}. Secondly, rather than using independent Cochran-Armitage tests for the resulting proportions as in \cite{Green2014} (cherry-picking style for the p-values), these cut-point-selected categories are considered multiple correlated proportions and are analyzed jointly \cite{Hothorn2016}. A third item is the use of multiple unrestricted Dunnett-type \cite{Dunnett1955} or order-restricted Williams-type \cite{Williams1971} contrast tests \cite{Hothorn2016}  but not trend tests solely for a linear dose-response relationship (note that the title of Armitage's paper is \textit{Tests for linear trends in proportions}). Fourthly, use permutation tests within the framework of conditional inference \cite{Hothorn2006} to achieve an appropriate test for small $n_i$ without possible variance in the control (and/ or lower dose) groups for multiple correlated endpoints- a permutative max(max-test)) \cite{Hothorn2021aa}. Finally, R-code should be available, using elementary R-code for categorization and the \Robject{library(coin)} \cite{coin2} for simultaneous inference. \\
To demonstrate this new approach, a more realistic data example was used, namely the scores of \textit{Gon Phenotype} in the dataset \textit{exampleHistData} within the \Robject{library(RSCABS)} for F1 males in week 8 for the full-sample size treatment groups 1 (control), 2,4,5.

\begin{figure}[htbp]
	\centering
		\includegraphics[width=0.50\textwidth]{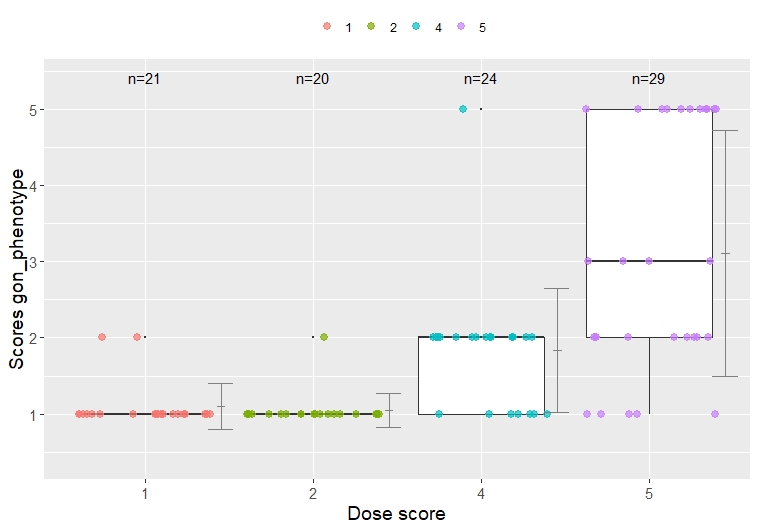}
	\caption{Boxplot with scores values}
	\label{fig:Plottall}
\end{figure}

The related R-code for a one-sided Dunnett-type test is for the three categories EP1, EP2, EP3, jointly:
\footnotesize
\begin{verbatim}
library(RSCABS)
data(exampleHistData)
subIndex<-which(exampleHistData$Generation=='F1' &
exampleHistData$Genotypic_Sex=='Male' &
exampleHistData$Age=='8_wk')
LH<-exampleHistData[subIndex, ]
lh<-LH[, c(2,6)]
lh$Gon<-as.numeric(lh$Gon_Phenotype)
lh$EP1<-ifelse(lh$Gon >1,1,0)
lh$EP2<-ifelse(lh$Gon >2,1,0)
lh$EP3<-ifelse(lh$Gon >3,1,0)
lh$treat<-as.factor(lh$Treatment)
lhh<-droplevels(lh[lh$treat!=6, ])
Lhh<-droplevels(lhh[lhh$treat!=3, ])

library("coin")
library("multcomp")
Co1 <- function(data) trafo(data, factor_trafo = function(x)
 model.matrix(~x - 1) %*% t(contrMat(table(x), "Dunnett")))
Codu <-independence_test(EP1 +EP2+EP3~ treat, data = Lhh, teststat = "maximum",
distribution = "approximate", xtrafo=Co1, alternative="greater")
pvalCODU <-pvalue(Codu, method="single-step")
pvalCODU
\end{verbatim}
\normalsize
The actual test is brief in the last eight lines: i+ii) packet calls, iii+iv) Dunnett-type contrast matrix, v+vi) multiple endpoint resampling max(max-test)), vii, viii) adjusted p-values for contrast and categories, namely:

\footnotesize
\begin{verbatim}
          Gon     EP1     EP2     EP3
2 - 1  0.9284  0.9463  0.9137  0.9216
4 - 1  0.2269  0.0001  0.8750  0.8570
5 - 1 <0.0001 <0.0001 <0.0001 <0.0001
\end{verbatim}
\normalsize

The interpretation of these adjusted p-values is intuitive: i) in principle the comparison of the high dose 5 vs. control always shows an increasing, strongly significant effect (globally seen independently of the cut-off),
ii) on the other hand the low dose (2) shows no effect at all (no matter which cut-off was considered), and iii) an increase is shown exactly only for cut-off 1 vs. (2, 3, 5); for the medium dose 4, small changes in severity $> 1$ already cause a significant effect - a toxicologically relevant finding. \\
This approach can be slightly modified for a Williams-type trend test and for simultaneous consideration of the score data itself in a Wilcoxon-type permutation test.

\footnotesize
\begin{verbatim}
CoW <- function(data) trafo(data, factor_trafo = function(x)
 model.matrix(~x - 1) %*% t(contrMat(table(x), "Williams")))
Cowi <-independence_test(EP1 +EP2+EP3~ treat, data = Lhh, teststat = "maximum",
distribution = "approximate", xtrafo=CoW, alternative="greater")
pvalCOWI <-pvalue(Cowi, method="single-step")
pvalCOWI
\end{verbatim}

\footnotesize
\begin{verbatim}
      EP1     EP2     EP3
C 1 <1e-04 <0.0001 <0.0001
C 2 <1e-04  0.0004  0.0036
C 3 <1e-04  0.0089  0.0386
\end{verbatim}
\normalsize

Monotone increasing trends exist for any category, whereas the strongest for 1, (2,3,5).

\footnotesize
\begin{verbatim}
Adu <-independence_test(Gon+EP1 +EP2+EP3~ treat, data = Lhh, teststat = "maximum",
distribution = "approximate", xtrafo=Co1, alternative="greater")
pvalADU <-pvalue(Adu, method="single-step")
pvalADU
\end{verbatim}

\begin{verbatim}
          Gon     EP1     EP2     EP3
2 - 1  0.9331  0.9517  0.9191  0.9264
4 - 1  0.2274 <0.0001  0.8821  0.8658
5 - 1 <0.0001 <0.0001 <0.0001 <0.0001
\end{verbatim}
\normalsize

A significant difference of the highest dose vs. control is also shown for the score data per se.
For the first data example the R-code is:
\footnotesize
\begin{verbatim}
library("coin")
library("multcomp")
green$treat<-as.factor(green$Dose)
Co1 <- function(data) trafo(data, factor_trafo = function(x)
 model.matrix(~x - 1) %*% t(contrMat(table(x), "Dunnett")))
gCodu <-independence_test(S12 +S23~ treat, data = green, teststat = "maximum",
distribution = "approximate", xtrafo=Co1, alternative="greater")
pvalGCODU <-pvalue(gCodu, method="single-step")
\end{verbatim}
\normalsize
The adjusted p-values  for the proportions S12 and S23 are:
\footnotesize
\begin{verbatim}
        S12    S23
2 - 1 0.9891 0.2772
3 - 1 0.9899 0.8290
4 - 1 0.9643 0.0707
5 - 1 0.2122 0.5731
\end{verbatim}
\normalsize

\subsection{Modified approach considering dose as a quantitative covariate}
Dose can be considered as qualitative factor using multiple contrast tests or as a quantitative covariate using linear or nonlinear dose-response models. A specific one-parametric slope fitting model approach, which allows a simplified interpretation and yet covers a wide range of dose-response models is Tukey's \cite{Tukey1985}, i.e., the maximum of untransformed, log-transformed, or ordinal-transformed dose scores \cite{Schaarschmidt2021}. This can also applied for multiple correlated proportions as asymptotic GLM-based models (but is only asymptotically valid) \cite{HRS2018}. Again, the R code is intuitive and relatively simple,  assuming dose values of 0,10,50, and 150 in the above example:

\footnotesize
\begin{verbatim}
library(tukeytrend)
library(multcomp)

Lhh$Dose<-as.numeric(as.character(Lhh$Treatment))
Lhh$dose[Lhh$Dose==1] <-0
Lhh$dose[Lhh$Dose==2] <-10
Lhh$dose[Lhh$Dose==4] <-50
Lhh$dose[Lhh$Dose==5] <-150

bx1<-glm(EP1~dose, data=Lhh, family=binomial(logit))
bx2<-glm(EP2~dose, data=Lhh, family=binomial(logit))
bx3<-glm(EP3~dose, data=Lhh, family=binomial(logit))
tx1 <- tukeytrendfit(bx1, dose="dose", scaling=c("ari", "ord", "arilog"))
tx2 <- tukeytrendfit(bx2, dose="dose", scaling=c("ari", "ord", "arilog"))
tx3 <- tukeytrendfit(bx3, dose="dose", scaling=c("ari", "ord", "arilog"))
tx <- combtt(tx1, tx2, tx3)
TX <- summary(asglht(tx))
@
\end{verbatim}
\normalsize 
A specific fitting one-parametric slope interpretation is already more complex: multiple models and multiple cut-points:

\footnotesize
\begin{verbatim}
Linear Hypotheses:
                                        Estimate Std. Error t value Pr(>|t|)    
tx1.glm.EP1.doseari: doseari == 0       0.025103   0.005065   4.956  < 0.001 ***
tx1.glm.EP1.doseord: doseord == 0       1.618118   0.312682   5.175  < 0.001 ***
tx1.glm.EP1.dosearilog: dosearilog == 0 1.444525   0.261639   5.521  < 0.001 ***
tx2.glm.EP2.doseari: doseari == 0       0.036360   0.009961   3.650  0.00143 ** 
tx2.glm.EP2.doseord: doseord == 0       3.278679   1.043687   3.141  0.00737 ** 
tx2.glm.EP2.dosearilog: dosearilog == 0 2.942857   0.966542   3.045  0.00936 ** 
tx3.glm.EP3.doseari: doseari == 0       0.031547   0.009878   3.194  0.00610 ** 
tx3.glm.EP3.doseord: doseord == 0       2.768226   1.019649   2.715  0.02431 *  
tx3.glm.EP3.dosearilog: dosearilog == 0 2.464308   0.943507   2.612  0.03207 *  
\end{verbatim}
\normalsize
The most sensitive is for the cut-point 1 vs 2,3,5 for the logarithmic dose score.

\section{Summary}
A new permutation test for analyzing histopathologic findings with severity based on their decomposition into multiple correlated proportions for multiple Dunnett or Williams-type contrasts is presented. The focus is on problem-adequate interpretation and use for small $n_i$ designs with possibly reduced variance in the control. An asymptotic variant for simple quasilinear models is also available, allowing a wide class of possible dose-response dependencies to be evaluated. Due to the use of the CRAN packages \Robject{coin, tukeytrend and multcomp}, the evaluation of real data is relatively easy.\\
The next future work is simulation for small $n_i$ designs with different data conditions up to zero-variance in the control.

\footnotesize

\end{document}